\documentclass{ws-procs9x6-cpt22}
\begin{document}

\newcommand{\refeq}[1]{(\ref{#1})}
\def\etal {{\it et al.}}

\title{Coherent states for trapped ions. Applications in\\
quantum optics and precision measurements.}

\author{B.\ Mihalcea}

\address{$^1$Natl. Inst. for Laser, Plasma and Radiation Phys., Low Temp. Plasma Dept.\\
077125 M\u agurele, Ilfov County, Romania}

\begin{abstract}
The evolution of squeezed coherent states (CSs) of motion for trapped ions is investigated by applying the time dependent variational principle (TDVP) for the Schr\"{o}dinger equation. The method is applied in case of Paul and combined traps, for which the classical Hamiltonian and equations of motion are derived. Hence, CS provide a natural framework to: (a) engineer quantum correlated states for trapped ions intended for ultraprecise measurements, (b) explore the mechanisms responsible for decoherence, and (c) investigate the quantum-classical transition.
\end{abstract}

\bodymatter

\section{Introduction}

The concept of dynamical symmetry was the outcome of the blending between group theory and quantum mechanics (QM), occurring in the years 1950. Quantum integrability involves rigorous investigation of dynamical symmetry. Investigations of nonclassical states of spin systems coupled to a harmonic oscillator (HO) enables one to explore the mechanisms responsible for decoherence and the quantum-classical transition. 

\section{Time dependent variational principle. Equations of motion}

The Time-Dependent Hartree-Fock (TDHF) phase space is the classical correspondent of the quantum space of states (i.e., a coherent states representation of the boson-mapped fermion space of states), so classical information gained from TDHF trajectories provides relevant info on the evolution in the quantum space of states \cite{Rowe10}. We consider the action integral \cite{Kra81} 
\begin{equation}\label{v2}
	S = \int \frac{1}{\left\langle \psi |\psi \right\rangle }{\left[\left\langle \psi |H|\psi \right\rangle - \Im{m}\left\langle \frac{\partial \psi} {\partial t}\Bigg|\psi \right\rangle \right]} \ , 
\end{equation}
where $\hbar H$ is the quantum Hamiltonian and $\psi $ is a vector in the Hilbert space ${\mathcal H}$, which belongs to the domain of the self-adjoint quasienergy operator $K(t)= H - i\partial/\partial t$ \cite{Ghe92}. By minimizing the action $\left(\delta S = 0\right)$ it results that the Schr\"odinger equation is rigorously obtained from the TDVP \cite{Kra81}. 

We apply the TDVP to a manifold ${\widehat {\mathcal M}}$ of test vectors, parameterized by the points of a finite $2n$ dimensional phase space ${\mathcal M}$. In case of elementary quantum systems with dynamical symmetry groups that admit CSs, the complex structure is global and $\widehat {\mathcal M}$ represents a K\"ahler manifold. We consider $z = \left( z_1,z_2,\ldots ,z_n\right) \in {\mathcal O}$ as a system of complex canonical local coordinates in ${\mathcal M}$, where ${\mathcal O}$ is an open set from ${\mathbb C}^n$, of dimension $2n$. We choose a family of vectors $\psi \left( z\right) \in {\widehat{\mathcal M}}$ with $z\in {\mathcal O}$, holomorphic in $z$, such as ${\partial \psi \left( z\right) }/{\partial z^*_i} = 0 \,,\ i=1,\ldots ,n\ $. We introduce the matrix of the symplectic structure on $\widehat{\mathcal M}$ as \cite{Kra81, Mih18}

\begin{equation}\label{w2}
	\Omega =\left( \frac{\partial ^{2}}{\partial z_{j}\partial z_k^*}\ln
	\left\langle \psi \left( z^{\ast }\right) |\psi \left( z\right)
	\right\rangle \;\right) _{1\,\leq \,j,\,k\,\leq \,n}\  \ , 
\end{equation}
together with the matrix $\Lambda = -\mathrm{i}\left( \Omega ^{\ast }\right) ^{-1}$. Furthermore, we also introduce the Poisson bracket for the $f$ and $g$ functions, smooth on ${\mathcal{M}}$

	\begin{equation}\label{w3}
			\left\{ f,g\right\} = i\sum_{j, k = 1}^{n}\left( \Lambda_{j, k}\frac{\partial f}{\partial z_j}\frac{\partial g}{\partial z_k^{\ast }} - \Lambda_{k,j}^{\ast }\frac{\partial g}{\partial z_j}\frac{\partial f}{\partial z_k^{\ast}}\right) .  
		\end{equation}
We consider the following action integral 

\begin{equation}
S = \int\limits_{t_1}^{t_2}\left[ \Im m \left(\sum\limits_{i = 1}^{n} \frac{dz_j^{\ast}}{dt} \frac{\partial }{\partial z_j}^{\ast}\right) \ln \left\langle \psi \left( z^{\ast}\right) |\psi \left( z\right) \right\rangle - H_{cl} \left( z, z^{\ast}\right) \right] dt ,  \label{w4}
\end{equation}
where

\begin{equation}\label{w5}
	H_{cl}\left( z,z^{\ast }\right) = \frac{\left\langle \psi \left( z^{\ast }\right) \left\vert H\right\vert \psi \left( z\right) \right\rangle }{\left\langle \psi \left( z^{\ast }\right) |\psi \left( z\right) \right\rangle }.  
\end{equation}
By applying the TDVP for the action integral $S$ on the manifold $\widehat{{\mathcal{M}}}$, we derive the classical Liouville equations of motion

\begin{equation}\label{v34}
	\frac{dz_j}{dt} = \left\{z, H_{cl} \right\} \, \  ,\,\, \frac{dz^*_j}{dt} = \left\{z^*, H_{cl}\right\} \, \   ,   
\end{equation}
where $\widehat{{\mathcal{M}}}$ is considered as a CS orbit, while $H_{cl}\left(z,z^*\right)$ stands for the expectation value of the quantum Hamiltonian in the state represented by $\psi (z) \in {\widehat{\mathcal M}}$. Hence, $H_{cl}$ is considered the classical Hamiltonian associated to the quantum Hamiltonian $H$, an operation called {\it {dequantization}} \cite{Abri05, Mih18}.

\section{Semiclassical dynamics for a trapped ion. Coherent states}

We consider the quantum Hamiltonian for a particle of mass $m$ and electric charge $q$, confined in a combined 3D quadrupole ion trap (QIT) that exhibits axial symmetry  \cite{Ghe92, Mih18}
\begin{equation}
	H_{2} = \frac 1{2m}\left( -i\hbar {\mathbf \nabla } - \frac q2{\mathbf B}\times {\mathbf r}\right)^2 + qA(t)\left(
	x^2 + y^2 - 2z^2\right) \ ,  \label{w9}
\end{equation}
where $\mathbf r = \left(x,y,z\right)$ is the position operator, and $\mathbf B = \left(0,0,B_0\right)$ denotes the constant axial magnetic field. For a Penning trap $A$ is a constant, while for dynamical (Paul) traps ($B_0 = 0$) it is a time periodic function $$A(t) = \left(r_0^2 + 2 z_0^2\right)^{-1}\left( U_0 + V_0\cos \Omega t\right)$$
where $r_0$ and $z_0$ denote the trap semiaxes, $\Omega$ is the RF voltage frequency, while $U_0$ and $V_0$ stand for the d.c. and RF trapping voltages, respectively. Because the Hamiltonian $H_{2}$ commutes with the axial angular momentum operator  $L_{z}$,  
we restrict the analysis to a subspace of the eigenvectors asso-\newline ciated to this operator, with fixed eigenvalue $\hbar l$, where $l$ is the orbital quantum number. The Hamiltonian reduced to this subspace is \cite{Ghe92, Mih18}
\begin{equation}
	H_{2l} = H^{(\mathrm{a})} + H^{(\mathrm{r})} - \frac{\hbar \omega_{c}}{2}l\ ,
	\label{w12}
\end{equation}
where the axial $H^{(\mathrm{a})}$ and radial $H^{\mathrm{r}}$ are  
\begin{equation}
	H^{(\mathrm{a})} = -\frac{\hbar ^{2}}{2m}\frac{\partial^{2}}{\partial \rho^{2}}+\frac{m}{2}\lambda _{\mathrm{a}}z^{2}\ ,  \label{w13}
\end{equation}%
\begin{equation}
	H^r = -\frac{\hbar^2}{2m}\left( \frac{\partial^2}{\partial \rho^2} + \frac 1{\rho }\frac{\partial }{\partial \rho} - \frac{l^2}{\rho^2}\right) + \frac{m}{2}\lambda_r\rho^2 , \ \rho = \sqrt{x^2 + y^2}  \label{w14}
\end{equation}%
and
\begin{equation}
	\lambda_{\mathrm{a}} = - \frac{4q}{m}A\ ,\quad \lambda_{\mathrm{r}} = \frac 14(\omega_c^2 - 2\lambda_{\mathrm{a}})\ ,\quad \omega_{c} = \frac qm B_0\ .  \label{w15}
\end{equation}
The generators of the Lie algebra are \cite{Ghe92} 
\begin{equation}
	K_{0,1}^{(a)} = \frac 14\left( \mp \frac {\partial^2}{\partial z^2} + z^2\right) , \ K_{0,1}^{(r)} = \frac 14\left[ \mp \frac{\partial^2}{\partial \rho^2} + \rho^2 \pm \left( l^2 - \frac 14 \right) \frac 1{\rho^2} \right] .  \label{w18}
\end{equation}

The operators $K_{0}^{(c)}$, $K_{1}^{(c)}$ and $K_{2}^{(c)} = i[K_{1}^{(c)}, K_{0}^{(c)}]$ satisfy the commutation relationships as shown in \cite{Ghe92, Mih18}. These operators generate the Lie algebra associated to the Lie group $\mathcal G_c$, obtained by means of an unitary irreducible representation (UIR) of the symplectic group $Sp(2,\mathbb R)$ of Bargmann indices $k_a = \frac 14, \frac 34$, and $k_r = \frac {l+1}2\ $ \cite{Ghe92}. The axial $H^{(a)}$ and radial $H^{(r)}$ Hamiltonians are linear combinations of the generators:
\begin{equation}
	H^{(\mathrm{c})}=\alpha_c K_0^c + \beta_c K_1^c , \  c = a, r  \label{w21}
\end{equation}
with 
\begin{equation}
	\alpha_c = m \lambda_c + \frac{\hbar^2}m\ ,\quad
	\beta_c = m\lambda_c - \frac{\hbar^2}m\ .
	\label{w22}
\end{equation}

The solutions of the Schr\"{o}dinger equation for $H_{2l}$ (eq. \ref{w12}) are  
\begin{equation}
	\Psi_{k_am_ak_rm_rl} = \frac 1{\sqrt{\rho }}\exp \left[ \mathrm{i}l\left( \theta + \frac{\omega_c}2 t\right) - i\varphi \right] \psi_{k_am_a}(z_a)\psi_{k_r m_r}(z_r)\ ,
	\label{w23}
\end{equation}
with $m_a, m_r \in \mathbb N$, while $z_a, z_r \in \mathbb C$ ($ \left\vert z_{a,r} \right\vert < 1$), and $\varphi = (k_a + m_a)\varphi_a + (k_r + m_r)\varphi_r, \varphi \in \mathbb R$ is the phase. $\psi_{k_am_a}(z_a)$ and $\psi_{k_rm_r}(z_r)$ stand for coherent symplectic vectors for the axial $\mathcal G_a$ and radial $\mathcal G_r$ dynamical groups.

The quantum Hamiltonian for a trapped ion confined in a combined nonideal 3D QIT with axial symmetry can be expressed as \cite{Mih18}
\begin{equation}
	H_{l}= H_{2l} + qA\left( t\right) P\left( \rho^2,z^2\right) ,  \label{w34}
\end{equation}
where the anharmonic part is defined as a polynomial

\begin{equation}
	P\left( \rho^2,z^2\right) =\sum\limits_{k\geq 2}c_{k}H_{2k}\left( \rho, z\right) \ , \ H_{2k}\left( \rho, z\right) = \sum\limits_{j=0}^k\frac{(2k)!\rho^{2j}z^{2k-2j}}{4^{j}(2k-2j)!\left( j!\right)^2}  \label{w35}
\end{equation}

We show the energy function associated to the quantum Hamiltonian $H_{l}$ is a classical one $\tilde{H}_{l}$, whose values are exactly the expectation values of the $H_{l}$ on the symplectic coherent states $\psi_{k_a0}(z_a)$ and $\psi_{k_r0}(z_r)$. By applying the TDVP we derive the following equations of motion within the unit disk $\left\vert z_c\right\vert <1$:
\begin{equation}
	\frac {dz_c}{dt} =\left\{ z_c, \tilde{H}_l\right\}_c\;,  \label{w46}
\end{equation}%
where $\left\{ z_c,\tilde{H}_l\right\}_c$ denotes the Poisson bracket, defined as

\begin{equation}
	\left\{ f,g\right\}_c = \frac{\left(1 - z_c z_c^{\ast }\right)^2}{2\mathrm{i}k_c}\left( \frac{\partial f}{\partial z_c^{\ast }}\frac{\partial g}{\partial z_c^{\ast}} - \frac{\partial f}{\partial z_c}\frac{\partial g}{\partial z_c}\right) \;.  \label{w47}
\end{equation} 

We introduce the complex variables $\xi_c, \eta_c, \ c = a, r$ 

\begin{equation}
	\xi_c = \frac{\left( 1 + z_c\right) \left( 1 + z_c^{\ast}\right)}{1 - z_c z_c^{\ast }} ,\;\; \eta_c = \frac{\left( 1 - z_c \right) \left( 1 - z_c^{\ast }\right)}{1 - z_c z_c^{\ast }}\ .  \label{w49}
\end{equation}
Then, we write the classical Hamiltonian (Husimi function) as

\begin{eqnarray}
	\tilde{H}_{l} = A_r\eta_r + A_a\eta_a + B_r\xi_r + B_a\xi_a + \left( C_{20}\xi_r^2 + C_{11}\xi_r\xi_a + C_{02}\xi_a^2\right) +  \nonumber \\
	 \left( D_{30}\xi_r^3 + D_{21}\xi_r^2 \xi_a + D_{12}\xi_r\xi_a^2 + D_{03}\xi_a^3\right) \ - \frac{\omega_c}2\hbar l. \label{w50}
\end{eqnarray}

\section{Conclusions}

Quasiclassical dynamics of trapped ions is described by applying the TDVP for the Schr\"{o}dinger equation on coherent state (CS) orbits, introduced as sub-manifolds of the space of quantum states. The method enables one to derive the Hamilton equations of motion on symplectic manifolds. The results are also valid for the CM motion in case of a system of identical ions. The classical Hamiltonian associated to the system is obtained as the expectation value of the quantum Hamiltonian on CS. Such formalism can be applied to 3D nonlinear quadrupole ion traps (QIT), for which the quantum Hamiltonians are derived.

\end{document}